\documentclass[debug]{rmaa}

\usepackage{paralist}
\usepackage{amsmath} 
\usepackage{amsfonts}
\usepackage{psfrag,color}
 \usepackage[margin=1cm,font=footnotesize,labelfont=bf]{caption} 

\title{Estimation of the Star Formation Rate  using Long-Gamma Ray Burst observed by Swift } 

\author{
  M. El\'ias,\altaffilmark{1} 
  O. M. Mart\'inez,\altaffilmark{1} }

\altaffiltext{1}{Benem\'erita Universidad Aut\'onoma de Puebla}

\shortauthor{El\'ias, Mart\'inez}
\shorttitle{Study of the SFR through Long-GRBs }

\fulladdresses{ 
\item El\'ias and Mart\'inez: Benem\'erita Universidad Aut\'onoma de Puebla,
4 Sur 104, Centro, 5013 Puebla, Pue.(Mau$\_$584@hotmail.com).
}

\listofauthors{M. El\'ias , O. M. Mat\'inez , }
\indexauthor{El\'ias, M.}
\indexauthor{Mart\'inez, O.}

\abstract{In this work we estimate the Star Formation Rate (SFR) through  333 Long-GRBs detected by Swift. This investigation is based on the empirical model proposed by \citet{Yuksel2008}, basically, the SFR is estimated using long-GRBs considering that they have an stellar origin based on the Collapsar model or the collapse of massive stars (Hypernova) $M>20 M_{\bigodot} $. The analysis starts with the study of  $\varepsilon (z)$  which accounts the long-GRBs production rate and it is parameterized by  $\varepsilon(z)=\varepsilon_{0}(1+z)^{\delta} $ where  $\varepsilon_{0}$ include the SFR absolute conversion to GRBs rate in a  luminosity range already defined and $\delta$ is a dynamical parameter which changes at different regions of redshift it accounts the SFR slope which is obtained by an analysis of linear regression over our Long-GRBs sample, the results obtained provide evidence that support our proposal to use Long-GRBs as tracers of  SFR.}

\resumen{En este trabajo se estima la Tasa de Formaci\'on Estelar (SFR)  mediante el an\'alisis de una muestra de 333   \textit{\textbf{Gamma Ray Bursts}}  (GRBs) Largos detectados por  Swift.  Este estudio se basa en el modelo emp\'irico propuesto por \citet{Yuksel2008},  b\'asicamente, la SFR se calcula utilizando GRBs Largos tomando en consideraci\'on que son originados seg\'un el modelo Collapsar o del colapso de estrellas masivas tipo hipernova ( $M>20 M_{\bigodot} $). El an\'alisis parte del estudio de $\varepsilon (z)$   que representa la tasa de  producci\'on de GRBs Largos, parametriz\'andolo de la forma $\varepsilon(z)=\varepsilon_{0}(1+z)^{\delta} $, donde $\varepsilon_{0}$  incluye la conversi\'on absoluta de la SFR a la tasa de GRB  en un rango de luminosidad de GRB dado y el \'indice $\delta$ es un par\'ametro din\'amico que cambia con z y representa la pendiente de la traza dejada por la  SFR. Los resultados  favorecen la propuesta usar a los GRBs Largos como  trazadores de la SFR.}

\addkeyword{galaxies: star formation}
\addkeyword{gamma-ray burst: general}
\addkeyword{stars: massive}

\begin{document}
\maketitle

\section{Introdution}
\label{sec:intro}

Gamma Ray Burst are related to extremely energetic explosions in far away galaxies (for reviews, see \citet{Wang2015}; \citet{Wei2017}; \citet{Petitjean2016}  ), based on the collapse model which proposes the formation of Long-GRBs by the collapse of rapidly rotating super massive star (e.g. Wolf-Rayet star $M>20 M_{\bigodot} $, for cosmological implications of GRBs see \citet{Wei2017})

 we can trace and prove the SFR \citep{Yuksel2008} \citep{Kistler2008} \citep{WangFY2013} related with this events. The study of SFR  through traditional tracers as continuous UV \citep{Cucciati2012}, \citep{Schenker2013}, \citep{Bouwens2014}, recombinacion lines of: H$\alpha$, Far Infrared \citep{Magnelli2013}, \citep{Gruppioni2013}, radio emition and X-ray, are inefficient at high redshift  ($z>4$) {\citep{Schneider2015}} due for their sensitivity to extinction for gas and dust and the universe expansion.

\begin{figure}[!t]
\includegraphics[width=.53\linewidth,height=4.5cm]{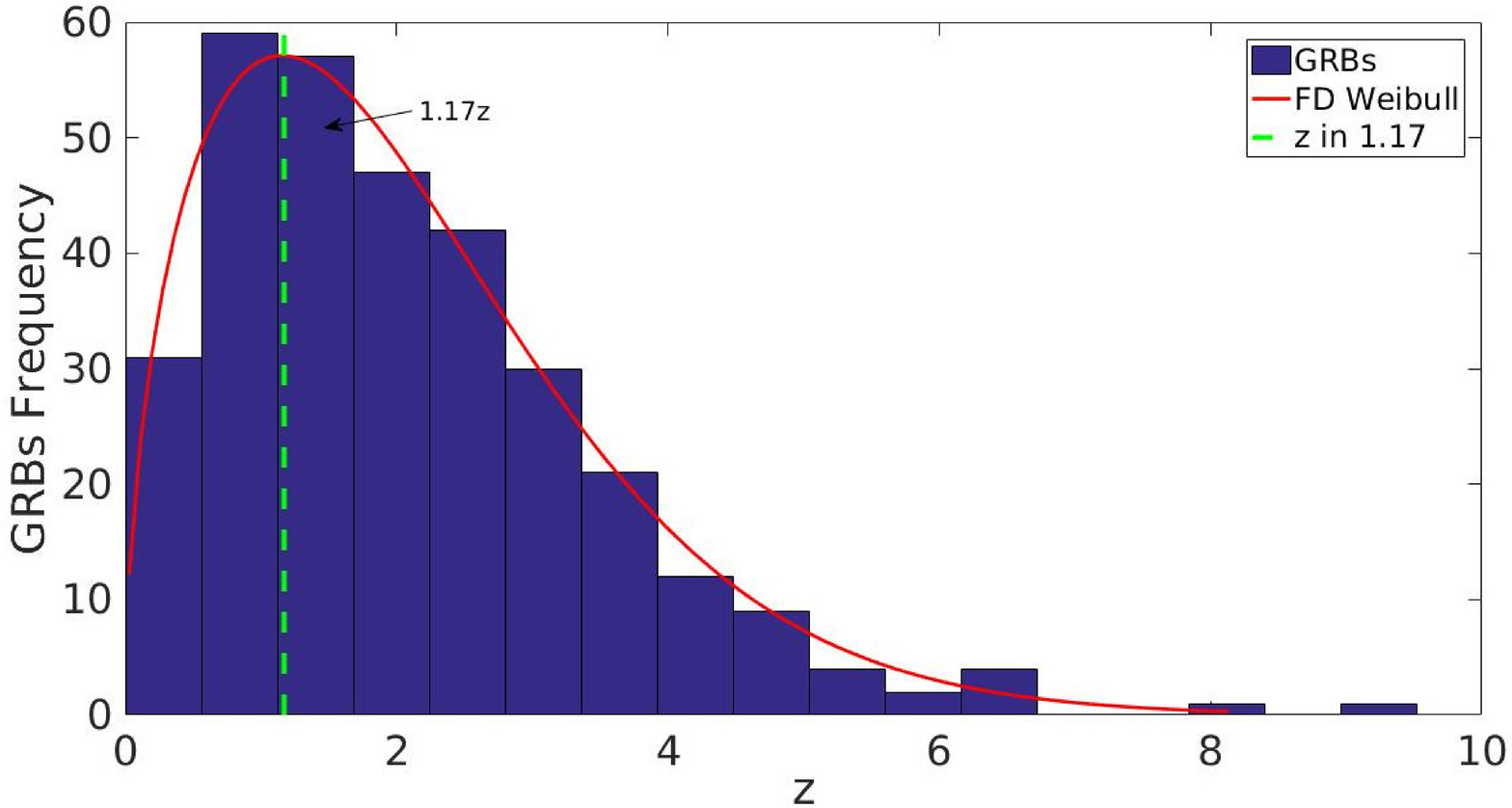} 
\includegraphics[width=.5\linewidth,height=4.5cm]{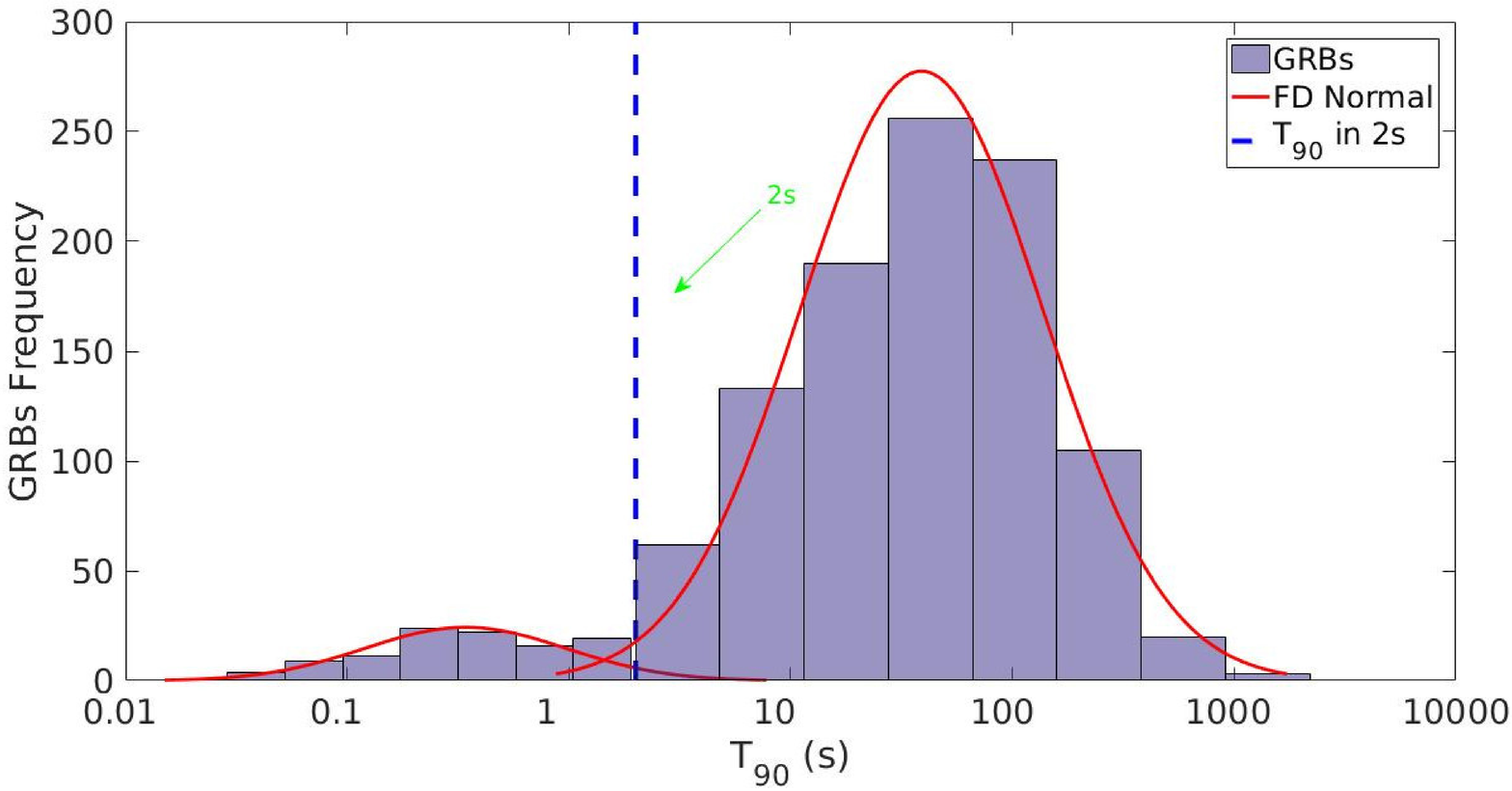} 
     \caption{ \textbf{Left} Frequency histogram of $333$ Long-GRBs  over z where the data  presents its mode at $1.17 z$ and mean at  $z =2.06$ these results match with  the  observational results. \textbf{Right} Bimodal distribution of our sample made up by   994 Long-GRBs where we can see both types of GRBs (Long $T_{90} >2s$ and Short $T_{90} <2s$)}.
     \label{z}
     \end{figure}

The stellar formation activity in the universe was very intense in the past, higher than now, in $z \sim 2.5$ about $10\%$ of all stars were formed and about $50\%$  of the local universe took place in $z\sim 1$, \citep{Schneider2015} , the star formation rate density is a function which evolves with the time, it has shown an increase of 10 times bigger between now and $z\sim 1$ holding until $z\sim 3\--4$ and finally it decreases at $z>4$ \citep{Hopkins2006}\citep{CarrollyOstile2007} \citep{Schneider2015} ,  the figure \ref{z}  show the distribution of our sample with the redshift,  where the data presents a  mode at   $z \approx 1.17$ and mean at $z \approx 2.06$  these results match with the observational results. 

The paper is organized as follows. In \S 2 we present the main properties of our Long-GRBs sample. In \S 3 we develop the mathematical model to calculate the SFR using Long-GRBs  as a tracers. In \S 4 we present the results based on the compute of $\delta $ obtained by an analysis of linear regression over the Long-GRB sample. We conclude in \S 5.

\section{Description of the sample}
\label{sec:errors}

The data sample used, includes 959 GRBs observed by Swift  supplied by \citet{Butler2007} and 35 bursts detected by FERMI from \citet{Singer2015},  BeppoSAX\@ from \citet{Frontera2009} and  ROTSE from \citet{Rykoff2009}  obtaining a total of 994 GRBs, where 333 are Long-GRBs  with $T_{90}$ and $z$ established, from these 333 only 263 presents an isotropical Energy $E_{iso}$ already defined. We consider bursts up to 2017 June 4, the fig \ref{z} shows the data considered, as it was observed by BATSE the bimodal distribution allow  to define the short and Long-GRBs. 
  
\section{Derivation of the SFR using GRBs}

The conversion factor  between  GRBs rate and SFR is hard to identify,  supported by an increasing amount of data of the cosmic star formation rate at low redshift  $z < 4 $ \citep{Cucciati2012}\citep{Dahlen2007}\citep{Magnelli2013} and the relationship between Long-GRB and star  formation based in the hypernova model  we can relate the observed GRBs in low redshift  with the SFR measurements considering an additional evolution of  GRBs rate with SFR \citep{Kistler2008}\citep{Yuksel2008}.
\\

GRBs distribution  per unit of redshift over all sky is giving by 
     \begin{equation}
\frac { \dot { dN }  }{ dz } =F(z)\frac { \varepsilon (z){ \dot { \rho  }  }_{ * }(z)\quad  }{ \left< { f }_{ beam } \right>  } \frac { \frac { { dV }_{ com } }{ { dz } }  }{ 1+z } 
      \label{dN}
     \end{equation}
Where  $0<F(z)<1$  accounts the probability to obtain the redshift related to afterglow from their host galaxy.  $\varepsilon (z)$ accounts the Long-GRBs rate production with additional evolution effects.  ${\left< { f }_{ beam } \right>}$ accounts the number of GRBs that are observed due for their beaming,  ${\dot { \rho  }  }_{ * }(z)$ accounts the SFR density where the dot represent comoving coordinate , $1/(1+z)$ is a factor related to cosmological time dilation.  $dV_{com} / dz$ \footnote{ the comoving volume is giving by $dV_{com} / dz=4\pi { D }_{ com }^{ 2 }* dD_{ com}/dz $
\\
 the comoving distance $dD_{com}$ is giving by $d{ D }_{ com }= c/{ H }_{ 0 } \int _{ 0 }^{ z }{ dz'({ \Omega _{ m }\left( 1+z' \right) ^{ 3 }+\Omega _{ \Lambda  }   })^{-1} }$} differential volume in comoving coordinates per redshift unit.  $\varepsilon (z)$ is parameterized  as $\varepsilon (z)= {\varepsilon}_{0} {(1+z)^{\delta}}$ where ${\varepsilon}_{0}$ is a constant which includes the absolute conversion from SFR to GRB  in a GRB luminosity range, $\delta$ accounts the  slope left by the trace of the SFR in a redshift range. 

    The table \ref{tablaa1} presents 10 elements of the sample, listing some spectral properties as \textbf {Energy Fluence\footnote{$\mathrm{(15\--150 \, keV)\,[erg\,cm^{-2}]}$},  \textbf{Peak Energy Flux}\footnote{$\mathrm{(15\--150 \, keV)\,[erg\,cm^{-2}\,s^{-1}]}$}, \textbf{Peak Energy} Flux\footnote{$\mathrm{(15\--150 \, keV)\,[ph\,cm^{-2}\,s^{-1}]}$} $E_{iso}$, \textbf{Ep} and \textbf{$T_{90}$}}, using $E_{iso}$ we can obtain  the Isotropical luminosity $L_{iso}$ by the equation \ref{Lis}
\begin{equation}
 { { L }_{ iso }=\frac { { E }_{ iso }(1+z) }{ { { T }_{ 90 } } }  }
\label{Lis}	
 \end{equation}

\begin{table}[!t]
\centering
\resizebox{\textwidth}{!}{%
\begin{tabular}{@{}ccccccccc@{}}
\toprule
N & GRB & z & $T_{90}$ & Ep [kev] & Energy Fluence & Peak Energy Flux & Peak Photon Flux & $E_{iso}$ [erg] \\ \midrule
1 & GRB140512A & 0.73 & 158.76 & 270.4481 & 1.29E-05 & 5.69E-07 & 7.09467 & 5.47E+50 \\
2 & GRB140518A & 4.71 & 61.32 & 46.5668 & 1.04E-06 & 5.38E-08 & 0.88978 & 4.98E+51 \\
3 & GRB141225A & 0.92 & 40.77 & 132.6695 & 2.59E-06 & 1.06E-07 & 1.27368 & 3.86E+51 \\
4 & GRB150301B & 1.52 & 13.23 & 106.8910 & 1.81E-06 & 2.14E-07 & 2.82063 & 1.14E+52 \\
5 & GRB150323A & 0.59 & 150.4 & 81.3815 & 5.40E-06 & 2.98E-07 & 4.42309 & 9.30E+49 \\
6 & GRB150403A & 2.06 & 38.28 & 227.8612 & 1.58E-05 & 1.48E-06 & 17.2206 & 3.07E+52 \\
7 & GRB150413A & 3.14 & 264.29 & 63.1096 & 4.50E-06 & 6.83E-08 & 0.986981 & 5.04e+51 \\
8 & GRB150818A & 0.28 & 134.39 & 74.8740 & 3.97E-06 & 1.12E-07 & 1.71705 & 3.31E+49 \\
9 & GRB150821A & 0.76 & 149.93 & 197.5467 & 2.18E-05 & 4.24E-07 & 5.02955 & 5.70E+51 \\
10 & GRB151029A & 1.42 & 9.28 & 31.3418 & 4.15E-07 & 8.87E-08 & 1.71218 & 9.01E+50 \\ \midrule
\end{tabular}%
}
\caption{Spectral properties of the sample.}
\label{tablaa1}
\end{table}

\begin{figure}[!h]
   \includegraphics[width=1\columnwidth]{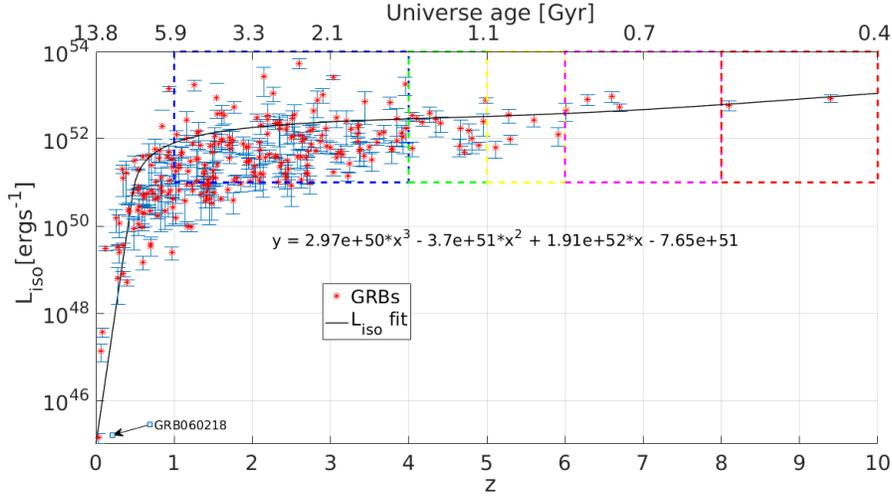}  
\caption{Distribution of 263 Long-GRBs detected by Swift from \citet{Butler2007}, we highlight 5 areas used to estimated the SFR density at different redshift bins $(1\--4, 4\--5, 5\--6, 6\--8, 8\--10)$ as discuss in the document, with $(173, 15, 4, 2)$ burst respectively.}
  \label{luminosidad}
\end{figure}

 In the figure \ref{luminosidad} we present the luminosity distribution of our sample made up by 263 Long-GRBs, here we observed the relation between $({L}_{iso})$ with redshift considering that only highly luminous  GRBs  can be seen in high z, using a luminosity boundary  of $\mathrm{{L}_{iso}>{10}^{51} erg {s}^{-1}}$  established by \citet{Kistler2008}, the spatial distribution of the events are in 5 redshifts bins $1\--4, 4\--5,  5\--6, 6\--8$ and $8\--10$ where we will calculate the SFR 
\\

The theoretical accounts of GRBs in the range of redshift from 1 to 4 are expressed by the equation \ref{N} \footnote{we use the values   ${\Omega}_{m}=0.3, {\Omega}_{\Lambda}=0.7$ based on the latest studies of Wilkinson Microwave Anisotropy Probe (WMAP) and Hubble Key Project (HKP) in a flat universe}. 
   \begin{equation*}
   { N }_{ 1-4 }^{ teo }=\Delta t\frac {  \Delta \Omega  }{ 4\pi  } \int _{ 1 }^{ 4 }{ dzF(z)\varepsilon (z)\frac { \dot { { \rho  }_{ * } } (z) }{ \left< { f }_{ beam } \right>  } \frac { \frac { { dV }_{ com } }{ dz }  }{ 1+z }  } 
     \end{equation*}
     
        \begin{equation}
        \label{N}
   { N }_{ 1-4 }^{ teo }=A\int _{ 1 }^{ 4 }{ dz\dot { { \rho  }_{ * } } (z){ (1+z) }^{ \delta  }\frac { \frac { { dV }_{ com } }{ dz }  }{ 1+z }  } 
    \end{equation}
Where 
\begin{equation*}
A=\frac {\Delta t \Delta \Omega F(z)\varepsilon_0 (z) }{ 4\pi \left< { f }_{ beam } \right>  } 
\end{equation*}
A depends in the total observed time by Swift $\Delta t$ and the angular sky coverage $\Delta \Omega$, utilizing the SFR overage density ${ \left< \dot { { \rho  }_{ * } }  \right>  }_{ { z }_{ 1 }-{ z }_{ 2 } }$  we compute the theoretical accounts of GRB in a range of redshift  from ${z}_{1}$ to ${z}_{2}$ is given by 

  \begin{equation}
  { N }_{ { z }_{ 1 }-{ z }_{ 2 } }^{ teo }={ \left< \dot { { \rho  }_{ * } }  \right>  }_{ { z }_{ 1 }-{ z }_{ 2 } }A\int _{ { z }_{ 1 } }^{ { z }_{ 2 } }{ dz{ (1+z) }^{ \delta  }\frac { \frac { { dV }_{ com } }{ dz }  }{ 1+z }  } 
  \end{equation}

  Taking  the calculus of  GRBs  observed ${ N }_{ { z }_{ 1 }-{ z }_{ 2 } }^{ obs }$ we obtain the SFR in a specific range of z,  ${ z }_{ 1 }\--{ z }_{ 2 }$  and using the  bin $ 1\--4$ we determine the SFR overage density  in the equation \ref{ecuacion8} 
  \begin{equation}
  \label{ecuacion8}
  { \left< \dot { { \rho  }_{ * } }  \right>  }_{ { z }_{ 1 }-{ z }_{ 2 } }=\frac { { N }_{ { z }_{ 1 }-{ z }_{ 2 } }^{ obs } }{ { N }_{ 1-4 }^{ obs } } \frac { \int _{ 1 }^{ 4 }{ dz\frac { \frac { { dV }_{ com } }{ dz }  }{ 1+z } { (1+z) }^{ \delta  } } \dot { { \rho  }_{ * } } (z) }{ \int _{ { z }_{ 1 } }^{ { z }_{ 2 } }{ dz\frac { \frac { { dV }_{ com } }{ dz }  }{ 1+z }  } { (1+z) }^{ \delta  } } 
  \end{equation}

\section{Description of the SFR model by Long-GRBs}

Considering the results obtained by \citet{Hopkins2006} and studies made by \citet{Yu2015} about the GRBs rate compared with SFR we defined the best fit to $\delta$ in different ranges of z,  where the best fit to $ \dot { { \rho  }_{ * } }$ is giving by the table  \ref{112}
        
 We calculate  $\dot { \rho  }_{ * } (z)$ parameterized as a function of redshift and $\delta$  using  a power law, considering that we are including a bigger range of redshift and also a bigger account of Long-GRBs than  \citet{Yuksel2008} we extend their  model with the equation \ref{bb} adding the term $\eta$ representing the overage account of Long-GRBs observed in the bin of z $(z_1,z_2)$ normalized by the  account of Long-GRBs in the bin (1,4).

\begin{equation}
\label{bb}
\dot { { \rho  }_{ * } } \left( z \right) =\eta \dot { { \rho  }_{ + } } \left( z \right) =\left( 1+\frac { { { N }_{ 1-4 }^{ obs } } }{ { { N }_{ { z }_{ 1 }-{ z }_{ 2 } }({ z }_{ 1 }+{ z }_{ 2 }) }/{ 2 } }  \right) \dot { { \rho  }_{ + } } \left( z \right)
\end{equation}     
  
 where $\dot { { \rho  }_{ + } } $ is given by equation \ref{ecuacion15}  proposed by \citet{Yuksel2008}, in order  not to  lose consistency we use $\dot { { \rho  }_{ + } } $ as $\dot { { \rho  }_{*} } $ .
  
\begin{equation}
 \label{ecuacion15} 
 \dot { { \rho }_{ * } } (z)=\dot { {\rho }_{ 0 } } { \left[{ \left( 1+z \right)  }^{ a\tau  }+{ \left( \frac { 1+z }{ B }  \right)  }^{ b\tau  }+{ \left( \frac { 1+z }{ C }  \right)  }^{ c\tau  } \right]  }^{ \frac {1}{\tau} }
 \end{equation}
  Where the constants $a, b$ y $c$   includes the logarithmic slope $\delta$ of the track left by $\dot { { \rho }_{ * } } (z)$ (see table \ref{112}), the normalization is $\dot { { \rho }_{ 0 } } =0.02M_{ \bigodot  }\,yr^{ -1 }\,Mpc^{ -3 }$ and $\tau \approx -10 $. ( see  \citet{Yuksel2008} for more details), we defined $A$ y $B$  with the next expressions

\begin{equation*}
B={ (1+{ z }_{ 1 }) }^{ 1-\frac { a }{ b }  }
 \end{equation*}
 \begin{equation*}
 C={ (1+z_{ 1 }) }^{ \frac { b-a }{ c }  }{ \left( 1+{ z }_{ 2 } \right)  }^{ 1-\frac { b }{ c }  }
 \end{equation*}
 
our first approximation of the density $\dot { { \rho }_{ * } } (z)$, using the best fit of $\delta$ from literature (see table \ref{112}) is

    \begin{equation}
    \label{ecuacion17}
 \dot { { \rho }_{ * } } (z)=0.02{ \left[ { \left( 1+z \right)  }^{ -30 }+{ \left( \frac { 1+z }{ 18.27 }  \right)  }^{ 9.4 }+{ \left( \frac { 1+z }{ 6.61 }  \right)  }^{ 43.6 } \right] }^{ -\frac { 1 }{ 10 }  }
 \end{equation}

\begin{center}
\begin{figure}[!t]
   \includegraphics[width=1\columnwidth]{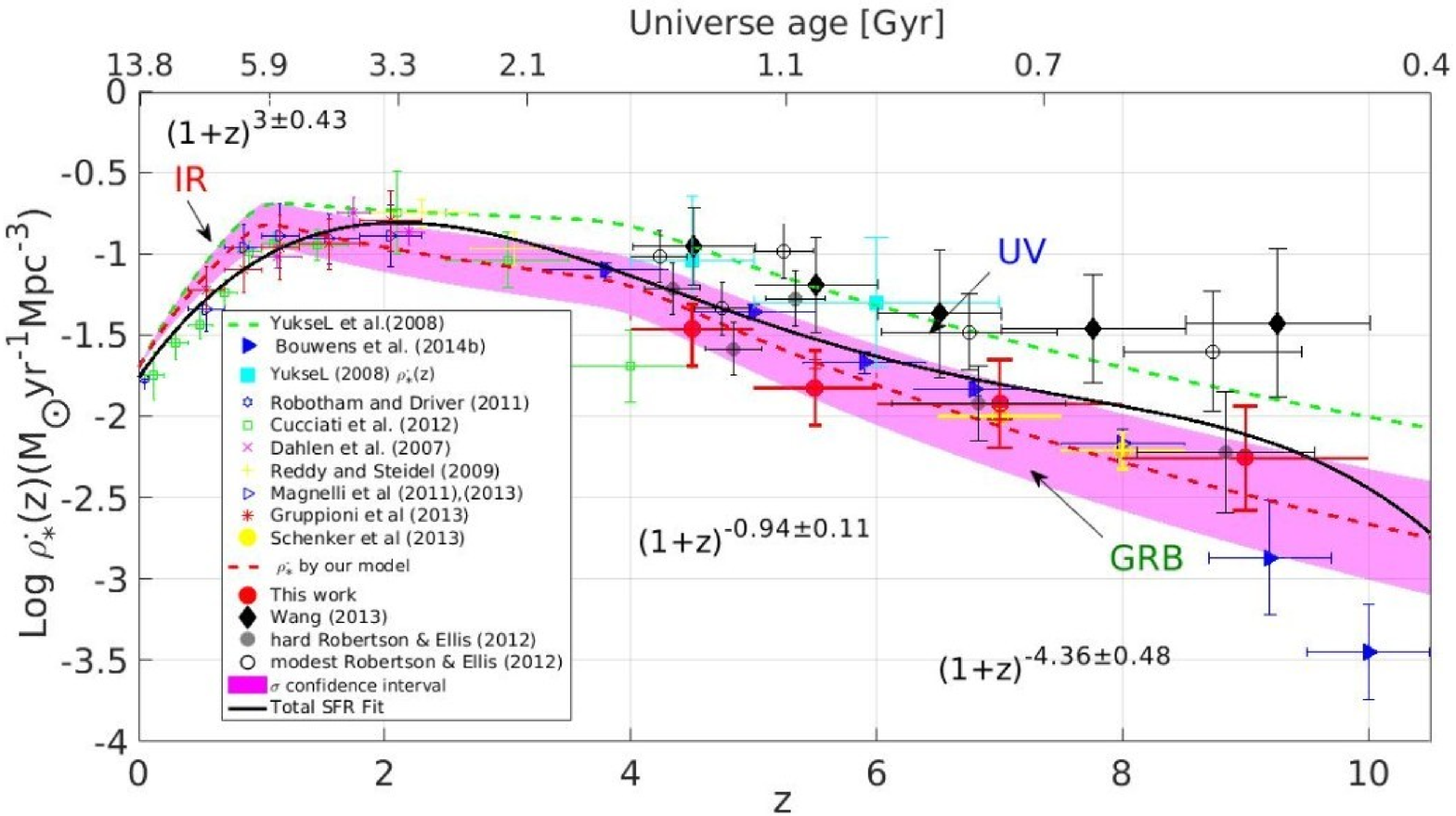} 
    \includegraphics[width=1\columnwidth]{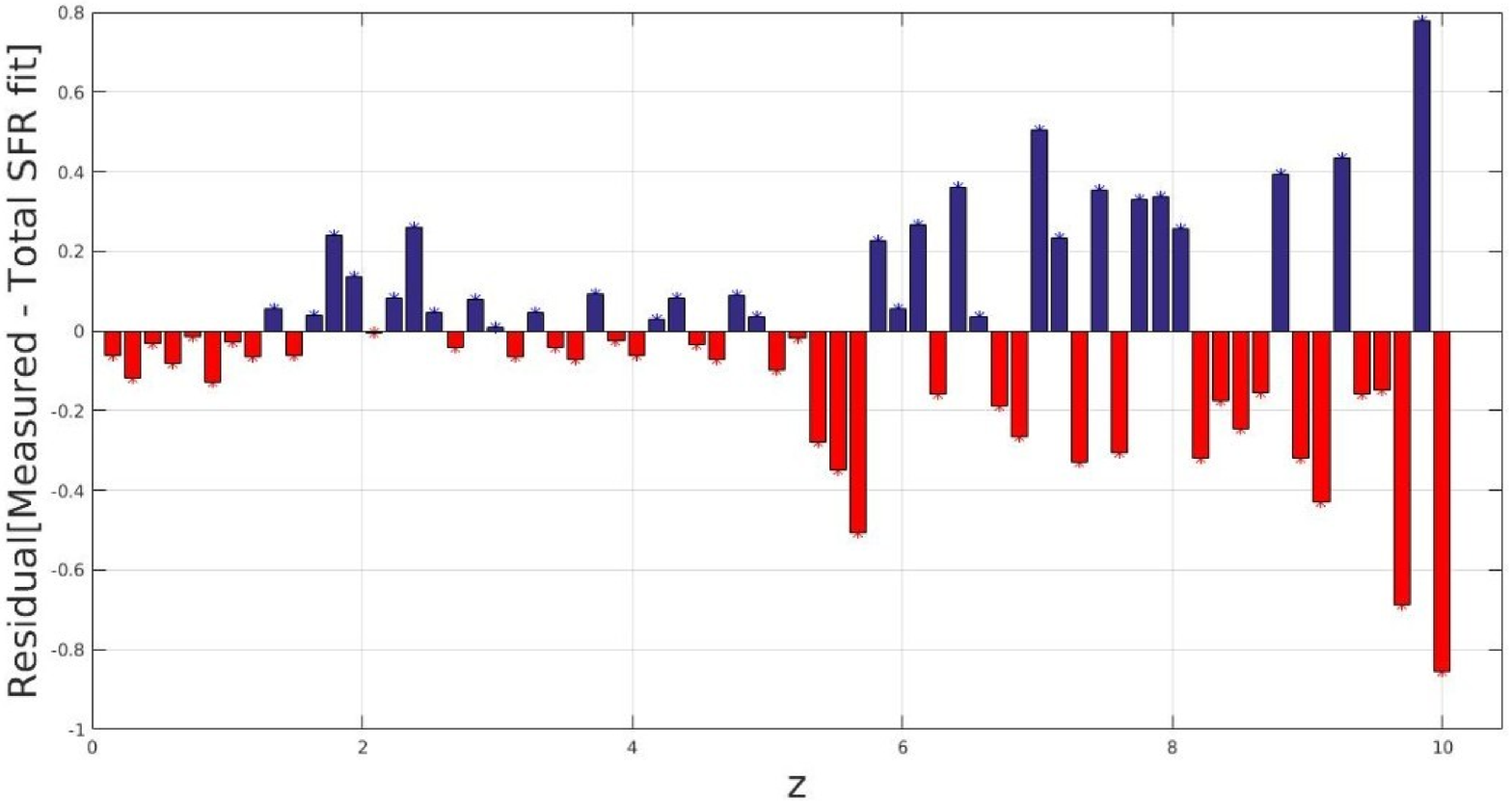} 
    \caption{ \textbf {Top}  Logarithmic distribution of $log (\dot{\rho}_* )(z)$   vs redshift where we present  the results and comparison between different tracers, the results are plotted in red solid diamonds, and results obtained in UV and FIR are plotted, the  wine color region accounts the confidence interval with $1\sigma $ of  significance and the black line represent the total SFR density fit with equation $log_{10}\,\dot{\rho}_* = -0.002z^4 + 0.053z^3 -0.414z^2 + 1.101z -1.764$ \textbf{Bottom}  Residual plot of  $log (\dot{\rho}_* )(z)$, we observe  significance high dispersion at high redshift $z>5$ symmetrical  distribution of the data show a good fit to the data. }
\label{SFR}
\end{figure} 
\end{center}

     In the figure \ref{SFR} it is shown the $\sigma$ confidence interval. The version update to the SFR in a specific range of $z$ of \citet{Yuksel2008}    used in this work is described by the  next equation.  

\begin{equation}
\label{nn}
{ \left< \dot { { \rho  }_{ * } }  \right>  }_{ { z }_{ 1 }-{ z }_{ 2 } }=\frac { { { N }_{ { z }_{ 1 }-{ z }_{ 2 } }^{ obs }+ }\frac { { { N }_{ 1-4 }^{ obs } } }{ \frac { { z }_{ 1 }+{ z }_{ 2 } }{ 2 }  }  }{ { N }_{ 1-4 }^{ obs } } \frac { \int _{ 1 }^{ 4 }{ dz\frac { \frac { { dV }_{ com } }{ dz }  }{ 1+z } { (1+z) }^{ \delta  } } \dot { { \rho  }_{ * } } (z) }{ \int _{ { z }_{ 1 } }^{ { z }_{ 2 } }{ dz\frac { \frac { { dV }_{ com } }{ dz }  }{ 1+z }  } { (1+z) }^{ \delta  } } 
  \end{equation}

\subsection{Statistical Analysis of the model}
       
 Considering  $\delta$ which accounts the  slope left by the trace of the SFR function in a redshift range is not constant  and taking account the relation between GRB with an stellar origin by the hypernova model \citep{Schneider2015}\citep{CarrollyOstile2007} we calculate these $\delta$s directly from the sample through linear regression over the z bins $0\--1  ,1\--4$ and $4\--10$ where every region has  $89, 214,$ and $30$ respectively and due that  z has 3 significant digits, we did the analysis using grouped data 
 
We calculate the frequency table of each bin and  their respective histogram, which lets us obtain the linear  regression over the data, getting their respectively slope, in the bin $0\--1$ with 89 burst we obtain the linear equation $y = 2.32x + 3.4286$, in the bin $1\--4$, with 214 burst  we obtain  the linear equation  $y = -1.0643x + 22.781$ and the bin $4\--10$, with 30 burst  we obtain  the linear equation  $y = -4x + 18$. proceeding with the analysis we calculate the confidence interval over one $\sigma$ of significance, getting the best fit to the model at different ranges of z,  this is shown in the table \ref{112}

  Based on the results of the statistical analysis we calculate the density $\dot { { \rho }_{ * } } (z)$ and the average density ${ \left< \dot { { \rho }_{ * } }  \right>  }_{ { z }_{ 1 }-{ z }_{ 2 } }$ ,in the figure \ref{SFR2} we compare the results  with the ones obtained by traditional tracers.

\section{Discussion and Conclusion}

\begin{center}
\begin{figure}[!h]
  \includegraphics[width=1\columnwidth]{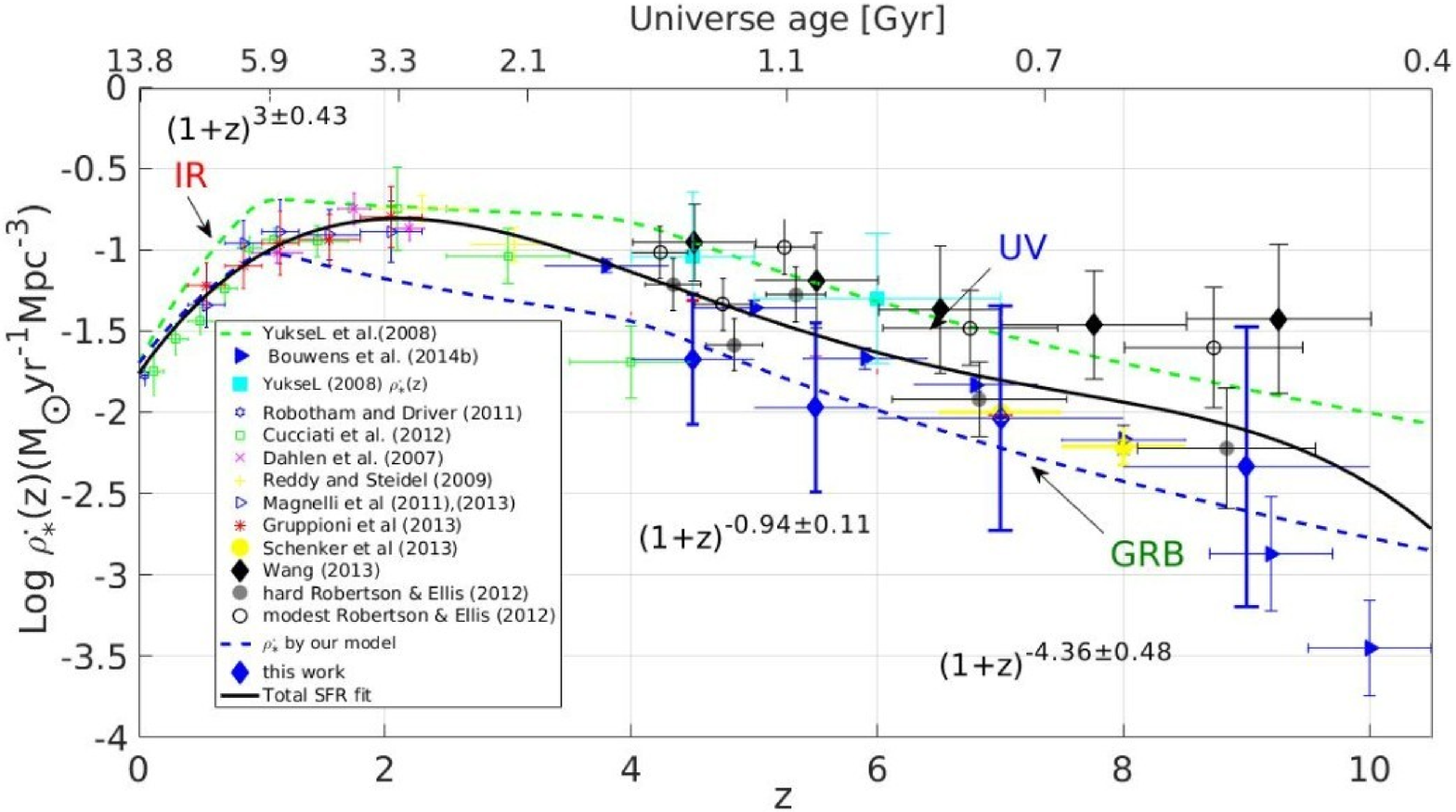} 
	    \caption{   Logarithmic distribution of SFR  $log (\dot{\rho}_* )(z)$   vs redshift analogous to the fig \ref{SFR} where we present  the results and comparison between different tracers, our results are plotted in blue solid diamonds using $\delta$ indexes from our statistical analysis and the black line represent the total SFR density fit }
\label{SFR2}
\end{figure} 
\end{center}

\begin{table}[!t]
\centering
\resizebox{12.5cm}{!}{%
\begin{tabular}{@{}ccccccccccc@{}}
\hline
Reference & Redshift range & $Log{ \left< \dot { { \rho}_{* } }\right>,} [M_{\bigodot}yr^{-1} Mpc^{-3}]$ & symbol in figure  \ref{SFR}, \ref{SFR2} \\ \hline
This work ($\delta$ proposed) & 4-5 & -1.47 & red solid diamond \\
 & 5-6 & -1.87 &  \\
 & 6-8 & -1.92 &  \\
 & 8-10 & -2.26 &  \\
This work ($\delta$ calculated) & 4-5 & -1.67 & blue solid diamond \\
 & 5-6 & -1.97 &  \\
 & 6-8 & -2.04 &  \\
 & 8-10 & -2.33 &  \\ \hline
\multicolumn{2}{c}{Redshift bins} & $\delta$  proposed & $\delta$ calculated \\ \toprule
0 & 1 & $3 \pm 0.43$ & $2.3 \pm 0.8$ \\
1 & 4 & $-0.94 \pm 0.11$ & $-1.1 \pm 0.2$ \\
4 & 10 & $-4.36 \pm 0.48$ & $-4 \pm 1.8$ \\ \midrule
\end{tabular}%
}
\caption{summarize between different values of the SFR obtained by this work.}
\label{112}
\end{table}

In this paper we presented the results of our work based in the estimation  of the SFR through a mathematical model which relates GRB directly with an stellar origin. we used the latest Swift catalog supplied by \citet{Butler2007}.  Based in the distribution of $L_{iso}$  (see figure \ref{luminosidad})  we computed the SFR using first  the values of $\delta$ from literature  (see figure \ref{SFR}), we made a linear regression analysis with our Long-GRB sample reproducing the reported $\delta$ indexes (see table \ref{112}),  using these results we compute a new values to SFR average density ${ \left< \dot { { \rho }_{ * } }  \right>  }_{ { z }_{ 1 }-{ z }_{ 2 } }$. We are including a bigger range of redshift than \citet{Yuksel2008} and  a bigger account of Long-GRBs than \citet{WangFY2013} we extend the model adding a new term $\eta$ (see equation \ref{nn}). our results are compared with the results from traditional tracers  as UV  and FIR (see figure \ref{SFR2}), in contrast to some other results such as \citet{Robertson2012} found higher and similar  values of $\dot{\rho}_*$ at $ z > 4 $ than ours based in a modest and hard evolution of the SFR with $\delta = 0.5$ and $\delta = 1.5$ respectively   considering GRBs from low metalicity host galaxies with  $12 + log[O/H] \approx 8.7$ \citep{Savaglio2005}  Their results with $\delta = 0.5$ and $\delta = 1.5$ are shown as open black circles  and solid gray circle  in  figure \ref{SFR} and \ref{SFR2} Our results can be marginally consistent with the gray circles. \citet{WangFY2013} used a sample of 110 luminous Swift GRBs to find an index value of $\delta \approx 0.5$ based on the origin of GRBs produced by rapidly rotating metal-poor stars with low mass, their SFR is higher  than our results. This may be  a consequence  for the update used of the Swift  GRB sample in our work and the type of model proposed  for the estimation of SFR considering our model highly dependence in the selected index value $\delta$ at different redshift bins

 considering  the physical implication and the results obtained along this work we conclude the next points.

\begin{itemize}
\item Considering the index $\delta$ represents the slope due for the SFR trace at different evolution stages of the universe, some previous studies have concluded that star formation dependents based on GRB at high redshift would be sufficient to maintain cosmic reionization over $6< z <9$  (e.g., \citet{Yuksel2008} ; \citet{Kistler2008}).  This possibility affect directly in the index value $\delta$ giving a minimums and maximus values for this parameter when  observational results show that GRBs are prompts to appear in low metallicity host galaxies \citep{Savaglio2005} implying a possible  metallicity limits for a massive star to transform into an successful GRB. Concluding that the decreasing of cosmic metallicity may to rise the relative number of GRBs  at high redshift and decrease to the local universe \citep{Butler2007} this observational results constrain the values of  $\delta$ obtained by our model using regression analyses over our GRBs Swift sample.

\item The figure \ref{z} the frequency histogram of frequency distribution of  333 Long-GRBs with redshift  show a Weibull distribution with mode at  z $\approx$ 1.17 and mean at z $\approx 2.06$. these values match with observational results of SFR , considering that in z $\sim$ 2.5, about 10$\%$ of all stars were formed and about 50$\%$  of the local universe took place  at  $z\sim 1$, \citet{Schneider2015}.

\item  We computed the values of the $log { \left< \dot { { \rho }_{ * } }  \right>  }_{ { z }_{ 1 }-{ z }_{ 2 } } $ using both values of $\delta$  from literature and  by our linear regression analysis where the best fit to $\delta$, it is shown in table \ref{112} our results match with the results from traditional tracers as UV, and FIR,  this  provide evidence that support our proposal to use Long-GRBs as tracers of SFR. 

\item  The Isotropically  luminosity  distribution  $L_{iso}$ (see figure \ref{luminosidad}) presents one particular outlier, the Long-GRB 060218 in $z = 0.03$ with the lowest $L_{iso}$  and also the largest $T_{90}$ ( $\approx 2100s$) this atypical  values  convert this event into  a new topic to investigate due for its strange properties.
\end{itemize}

\citep{Nicastro2018}

\end{document}